\documentclass[12pt]{iopart}

\usepackage{iopams}  

\usepackage{bm}
\usepackage{graphics} 
\usepackage{graphicx}  
\usepackage{amsfonts}  
\usepackage{amssymb}
\usepackage{amsthm}
\usepackage{mathrsfs}
\usepackage{subfigure}
\usepackage{color}

\renewcommand{\phi}{\varphi}
\renewcommand{\>}{\right \rangle}
\newcommand{\<}{\left \langle}
\newcommand{\ket}[1]{\left |#1\>}
\newcommand{\bra}[1]{\<#1\right |}

\newcommand{\be}{\begin{equation}}
\newcommand{\ee}{\end{equation}}
\newcommand{\bea}{\begin{eqnarray}}
\newcommand{\eea}{\end{eqnarray}}

\begin{document}
\title[REMPI of Neon under XUV FEL radiation: A case study of the role of  harmonics]{Resonantly Enhanced Multiphoton Ionization under XUV FEL radiation: A case study of the role of  harmonics}

\author{G M Nikolopoulos$^1$ and P Lambropoulos$^{1,2}$}

\address{$^1$Institute of Electronic Structure \& Laser, FORTH, P.O. Box 1385, GR-70013 Heraklion, Greece}
\address{$^2$Department of Physics, University of Crete, P.O. Box 2208, GR-71003 Heraklion, Crete, Greece}

\date{\today}

\ead{nikolg@iesl.forth.gr}

\begin{abstract}

We provide a detailed quantitative study of the possible
role of a small admixture of harmonics on resonant two-photon ionization. 
The motivation comes from the occasional
presence of 2nd and 3rd harmonics in FEL radiation. We
obtain the dependence of ionic yields on the intensity
of the fundamental, the percentage of 2nd harmonic and the
detuning of the fundamental from resonance. Having
examined the cases of one and two intermediate resonances,
we arrive at results of general validity and global
behavior, showing that even a small amount of harmonic
may seem deceptively innocuous.

\end{abstract}

%\pacs{32.80.Aa, 32.80.Hd, 32.70.Jz, 32.80.Rm}

\submitto{\JPB: Special Issue on Frontiers of FEL Science II}

\maketitle

\section{Introduction}

The radiation of the new XUV to X-ray FEL (Free Electron Laser) sources is known to contain a small component of the 2nd and 3rd harmonic of the photon energy chosen for a
particular experiment \cite{Ack07,Emm10,SalSchYur98,SalSchYur03,SalSchYur10,Kri06,Kri03}. Although, depending on the specifics of the source the intensity of those components may vary, typically they amount to a few percent \cite{remark}. Again, depending on the particular source and experimental set up, various types of filtering, may reduce their intensity to much lower percentages of the fundamental. Be that as it may, if the process under investigation relies on single-photon ionization by the fundamental, then the ionization yield due to a harmonic of intensity, say 2\% of the fundamental, will be roughly 2\% of the yield due to the fundamental. The relative amounts of ionization may of course depart from the direct analogy between the relative intensities, owing to some differences in the respective cross sections, which in general do depend on the photon energy.

The situation changes drastically, however, if the focus of the investigation involves a non-linear process induced by the fundamental. Assume, for example, that the aim is to observe ionization yields due to non-resonant 2-photon absorption, within the regime of validity of LOPT (Lowest (non-vanishing) Order of Perturbation Theory), which is typically valid for the peak intensities and pulse durations presently available, in the XUV and beyond. In that case, the presence of the 2nd harmonic would also produce ions, through single-photon absorption. As is well known, the latter is proportional to the photon flux, while the non-resonant 2-photon process is proportional to
the square of the flux \cite{LNPL06}. Obviously, at sufficiently low intensity, the linear process may dominate, even if the amount of the 2nd harmonic is only a few percent of the fundamental. Given that the sources under consideration are pulsed, the intensities change with time. Thus, although during the rise and fall of the pulse, the linear process will dominate, near and around the peak, the 2-photon process may or may not take over; depending on the precise magnitude of the peak intensity.

Clearly, the above discussion cannot provide even a qualitative assessment, but it does point to the need for time-dependent modelling that includes the basic features of a pulse. There has actually been a relevant detailed quantitative study in the literature \cite{LNPL06},  dating back to 2006, addressing precisely the role of the harmonics in non-resonant 2-photon ionization of Helium under FEL radiation of photon energy 13 eV. Although the connection to one of the early experiments \cite{Laar05} at the first version of FLASH turned out to be somewhat uncertain, the theory did nevertheless demonstrate that even an amount of harmonic as low as 0.1\% of the fundamental can have a profound effect on the expected behaviour of the ion signal. Since under non-resonant 2-photon ionization, the laser intensity dependence of the ion signal should be proportional to the square of the intensity, the presence of even as small an amount of 2nd harmonic as the above is found to alter that power dependence significantly, masking thus the basic signature of the desired process. The interested reader will find in Ref. \cite{LNPL06} a number of further details illustrating the interplay between the fundamental and the 2nd and 3rd harmonics.

Whereas non-resonant N-photon ionization displays an unequivocal signature of a power law, with the ion yield being proportional to the Nth power of the intensity, the presence of an intermediate resonant state introduces a host of additional effects which preclude that simple power law dependence. Consequently, it is not a priori obvious how the ion signal would be modified by the presence of harmonics and how would one evaluate their impact on the process.  This question, having arisen recently in experiments at the FEL FERMI in connection with resonant or near-resonant 2-photon ionization of Neon \cite{private}, has provided the motivation for the present work.
The field of REMPI (Resonantly Enhanced Multiphoton Ionization) has a history of more than 40 years and represents a valuable tool of laser spectroscopy with applications to fundamental as well as applied physics and chemistry \cite{ShoreBook,AgoJPB78,PLLyrasPRA89}. The most general case of REMPI would be an N-photon transition from the ground to a bound (discrete) state which is connected to the continuum by an M-photon process, referred to as N+M REMPI, or alternatively as N-photon resonant (N+M)-photon ionization. The cases of N=M=1 and (N=2, M=1) are the most common and useful in practice. For our purposes in this work, the case N=M=1, also known as resonant 2-photon ionization, contains all of the essential physics needed for the elucidation of the motivating question in connection with the role of the harmonics.  

Although the overall process of 2-photon REMPI involves the absorption of 2 photons, depending on the laser intensities and bandwidth, coupling matrix elements and pulse durations, the ionic signal at the end of the pulse more often than not will not exhibit a simple power law dependence on the peak intensity. The coupling of the two resonant discrete states is proportional to the field amplitude, while the ionization from the excited state is proportional to the intensity. In terms of quantum optics language, we have a two-level system coupled to a continuum; an open quantum system.  A particular combination of intensity and bandwidth may lead to Rabi oscillations between the resonant discrete states, in which case a simple power law dependence on the intensity cannot be expected. In short, the overall process cannot be described by a single transition rate through Fermi’s golden rule, in terms of a two-photon cross section, as in the non-resonant case. The quantitative description requires a formulation in terms of the density matrix which can also incorporate all necessary features and parameters; such as temporal pulse shape, stochastic bandwidth if relevant and of course the detailed evolution of the system during the pulse, through which the underlying physical processes can be assessed.

In any process involving photoionization, the determination of ionic yields represents the simplest and least demanding measurement and this is our concern in this paper. As we will show, even within this narrow set of measurements, important and useful insight can be gained about the role of the harmonics. We do want to point out at the outset, however, that more refined measurements, such as photoelectron energy and possibly angular distribution spectra, can help to disentangle the contribution of the harmonics from that of the fundamental. To start with, the energy of the photoelectrons due to the 3rd harmonic can be distinguished from that due to the fundamental. On the other hand, the energies of the photoelectrons due to the 2nd harmonic coincide with those due to the 2-photon process by the fundamental. Further refinement through the additional measurement of angular distributions may help in distinguishing between the two, but only to a limited extent. It is reasonable to argue that the features imprinted on the ionic yields have more general validity, while those obtained through angular distributions would be strongly system dependent.
   
After a brief description of the system in section II, section III provides the detailed formal framework for the problem. Section IV contains the bulk of the numerical
results, with the detailed discussion of the interplay of
the various parameters that determine the final outcome. A
summary with concluding remarks is given in section V.

\begin{figure}
\begin{center}
\includegraphics[scale=0.7]{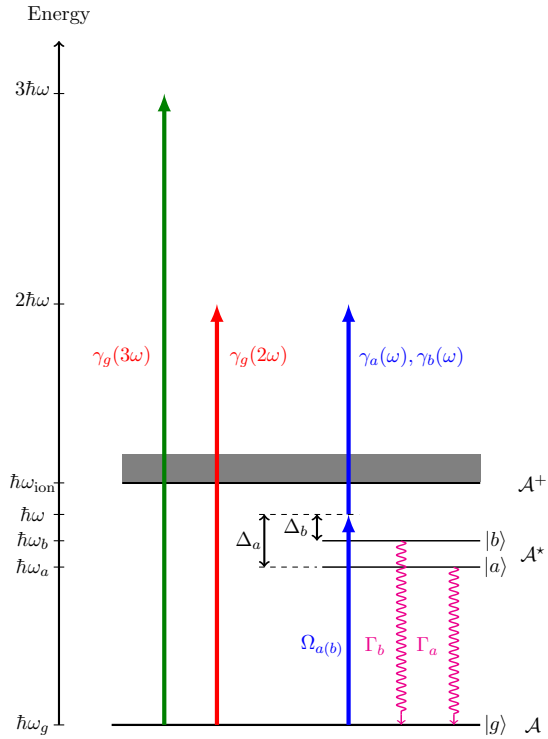}% Here is how to import EPS art
\caption{\label{IonScheme} Schematic representation of the system under consideration. The neutral atom ${\cal A}$ is ionized directly by absorbing single photons at the second and third harmonic, or two photons 
at the fundamental frequency, which is close to intermediate resonances (excited atom denoted by ${\cal A}^\star$ ). The corresponding ionization rates are denoted by $\gamma_g(2\omega)$, 
$\gamma_g(3\omega)$ and $\gamma_{a(b)}(\omega)$, respectively. The intermediate resonances 
are driven by the FEL radiation at frequency $\omega$, with  Rabi frequencies $\Omega_{a(b)}$ and  detunings $\Delta_{a(b)}$. Spontaneous emission at rate $\Gamma_{a(b)}$ is in principle present, but it is negligible for the time scales of interest.}
\end{center}
\end{figure}

\section{The system}
In a rather general context, the system under consideration is depicted in Fig. \ref{IonScheme}. A neutral atom interacts with FEL pulses at frequency $\omega$, which is close to one or two resonances (with transition frequencies $\omega_a - \omega_g$ and $\omega_b-\omega_g$), giving thus rise to REMPI. 
The radiation produced in typical FEL facilities (such as FERMI), besides the fundamental frequency $\omega$, also includes 
its harmonics $2\omega$ and $3\omega$. 
The presence of the 2nd harmonic is a spurious effect that is attributed to imperfections 
in the FEL mechanism (e.g., in the undulator), whereas the presence of the 3nd harmonic is a natural effect \cite{remark}. Although the harmonics' intensities, in general, are a very small
fraction of the intensity at the fundamental frequency
(typically $\lesssim 1\%$ each), they do give rise to direct single-photon ionization. Our objective here is to explore whether and under what conditions the ionic signal induced by the harmonics may become comparable to that of the REMPI.

The unperturbed atomic Hamiltonian is denoted by $\hat{\mathscr{ H}}_0$, with $\hat{\mathscr{ H}}_0\ket{\eta}=\hbar\omega_\eta\ket{\eta}$ and $\eta\in\{a,b\}$,  whereas the interaction between the field and the atoms is 
\bea
\hat{\mathscr{V}}(t)=-\hat{{\bm \mu}}\cdot \vec{\bf E}(t), 
\eea
where $\vec{\bf E}(t)$ is the time-dependent field evaluated at the position of the nucleus, and $\hat{{\bm \mu}}=e{\bm r}$ is the electric dipole operator. In the following we assume a field linearly polarized along the z direction and propagating along the x direction, with a time varying amplitude $E(t)$. Thus, the interaction term reduces to
\bea
\hat{\mathscr{V}}(t) = -eE(t)\hat{z},
\eea
with the total electric field given by 
\[
E(t) = E_{\omega}(t)+ E_{2\omega}(t)+E_{3\omega}(t)\]
where 
\bea
E_{q\omega}(t) \equiv {\cal E}_{q\omega}(t)e^{i q\omega t}+{\cal E}_{q\omega}^\star (t)e^{-i q\omega t}
\label{Eqomega}
\eea
with $ E_{q\omega}(t)$  denoting the electric field at the  carrier frequency  $q\omega$.

\section{Equations of Motion}
The reduced atomic density matrix $\rho(t)$  obeys the equation of motion 
$i\hbar \dot{\rho} =[\hat{\mathscr{ H}}_0+\hat{\mathscr{V}}(t),\rho]$.
Introducing the slowly varying amplitudes 
$\rho_{jj}=\sigma_{jj},\,\rho_{g\eta}=\sigma_{g\eta}e^{i\omega t},\, \rho_{ab}=\sigma_{ab}$,
applying the rotating-wave approximation (RWA) and introducing the decay channels, we have 
\bea
\frac{\partial\sigma_{gg}}{\partial t}&=&\Gamma_{a} \sigma_{aa}+ \Gamma_{b} \sigma_{bb} -\gamma_g(2\omega) \sigma_{gg} -\gamma_g(3\omega)\sigma_{gg}
\nonumber\\
&&+
2{\rm Im} \left [ \Omega_{g,a}^{\star}\sigma_{ga}\right ] + 
2{\rm Im} \left [ \Omega_{g,b}^{\star}\sigma_{gb}\right ] 
\nonumber\\
\frac{\partial\sigma_{\eta\eta}}{\partial t}&=&-[\gamma_{\eta}(\omega)+\Gamma_{\eta}]\sigma_{\eta\eta}
-2{\rm Im} \left [ \Omega_{g,a}^{\star}\sigma_{g\eta}\right ],\quad 
\eta\in\{a,b\} 
%%+i \left (\Omega_D^\star\sigma_{01}-\sigma_{01}^\star \Omega_D\right )
%\frac{\partial\sigma_{aa}}{\partial t}&=&-(\gamma_{a}+\Gamma_{a}) \sigma_{aa}
%-2{\rm Im} \left [ \Omega_{g,a}^{\star}\sigma_{ga}\right ] 
%%+i \left (\Omega_D^\star\sigma_{01}-\sigma_{01}^\star \Omega_D\right )
%\nonumber\\
%\frac{\partial\sigma_{bb}}{\partial t}&=&-(\gamma_{b}+\Gamma_{b}) \sigma_{bb}
%-2{\rm Im} \left [ \Omega_{g,b}^{\star}\sigma_{gb}\right ] 
\nonumber\\
\frac{\partial\sigma_{ga}}{\partial t}&=& 
\left ( i\Delta_{a} -\frac{\Gamma_{ga}}2\right )\sigma_{ga}
+i\left [ \Omega_{g,a} (\sigma_{aa}-\sigma_{gg})+\Omega_{g,b}\sigma_{ab}^\star\right ]
\nonumber\\
\frac{\partial\sigma_{gb}}{\partial t}&=&\left (i\Delta_{b} -\frac{\Gamma_{gb} }2
\right )
\sigma_{gb}
+i\left [ \Omega_{g,b}(\sigma_{bb}-\sigma_{gg}) + \Omega_{g,a}\sigma_{ab}\right ]
\nonumber\\
\frac{\partial\sigma_{ab}}{\partial t}&=& 
\left ( i\Delta_{ba} -\frac{\Gamma_{ab}}2 \right )\sigma_{ab}-i\Omega_{g,b}\sigma_{ga}^\star+i\Omega_{g,a}^\star\sigma_{gb}
\nonumber
\eea
where $\Delta_{\eta} = (\omega_{\eta}-\omega_g) -\omega$,  
$\Delta_{ba} = \omega_b-\omega_a$,
$\Gamma_{g\eta} = \Gamma_{a}+\gamma_{\eta}(\omega)+\gamma_g(2\omega)+\gamma_g(3\omega)$,  and $\Gamma_{ab} = \Gamma_{a}+\Gamma_{b}+\gamma_{a}(\omega)+\gamma_{b}(\omega)$. 

The  ionization rate at frequency $q\omega$  is time
dependent, given by  
\bea
\gamma_j(q\omega;t)= \sigma_j^{(1)}(q\omega) F_q(t)
\label{gamma_j}
\eea
where $\sigma_j^{(1)}$ is the corresponding single-photon cross-section, while  the flux of photons at the particular frequency  (in number of photons per ${\rm cm}^2$ per second) is given by 
\bea
F_q(t) = \frac{0.624}{q\hbar\omega[eV]}\times10^{19}\times I_q[W/cm^2].
\label{F_q}
\eea
To suppress notation the time dependence of $\gamma_\eta$ and $\gamma_g$ is not 
shown in the above equations of motion. 
From now on the intensity at the fundamental frequency $\omega$ is denoted 
by  $I(t) = I_1(t)$, whereas $I_q(t)$ refer to the $q$th harmonic. 
The intensity can be written as  $I(t) = I^{(0)} f(t)$, where $f(t)$ is the 
pulse profile and $I^{(0)}$ is the peak intensity. The intensities of the harmonics are fractions of  
$I(t)$ and can be expressed as $I_q(t) = r_q I(t)$. 

The Rabi frequency between two atomic states $\ket{g}$ and $\ket{\eta}$ 
is defined as 
\bea
\Omega_{\eta}(t)  &\equiv&  \frac{\bra{\eta}\hat{\mu}\ket{g}}{\hbar} E_\omega(t;0)  
\approx 2.207 \times 10^8  \mu_{\eta g} \times  \sqrt{I(t)},
\label{RabiEq}
\eea
where in the last expression the intensity  is measured in W/cm$^2$ and the dipole matrix element is in atomic units, and are typically estimated  through standard numerical codes and techniques.

For the sake of formal completeness, our model also  includes possible spontaneous decay channels.  The corresponding  rates, however, are of the order of $10^{-8}~$fs$^{-1}$ and
for pulses of duration up to a few hundred of femtoseconds, their effects can be safely  ignored. 

The most detailed monitoring of the ionization of the atom
through the various channels depicted in Fig. \ref{IonScheme} would be
obtained through the energy and angular distributions of
the emitted electrons. Formally, the photoelectron yields
for the four different ionization channels obey the
following equations of motion 
\bea
\frac{\partial Y^{(j)}_{q \omega}(t)}{\partial t} = \gamma_{j}(q\omega;t)\sigma_{jj};\,j\in\{g,a,b\}.
\label{yields:eom}
\eea
The total yields at the end of a pulse for the fundamental frequency and the harmonics 
are obtained by time integration of the equations of motion for the atomic density operator 
and the yields. 

The bandwidth of a pulse with fluctuations is given by  \cite{NikLam12-13}
\bea
\Delta\omega=\sqrt{\Delta\omega_{\min}^2+\Delta\omega_{\rm f}^2},
\label{Dw:eq}
\eea
where $\Delta\omega_{\rm f}$ is the bandwidth due to fluctuations which for Gaussian correlated noise is 
given by \cite{NikLam12-13}
\bea
\Delta\omega_{\rm f} = \frac{2\sqrt{2\ln(2)\pi}}{T_{\rm c}} \approx \frac{4.174}{T_{\rm c}}
\label{Tc:eq}
\eea
with $T_{\rm c}$ the coherence time of the source.  The bandwidth $\Delta\omega_{\min}$ is the 
Fourier-limited bandwidth, which for a Gaussian pulse with FWHM $\Delta t$ is given by 
\bea
\Delta\omega_{\min} = \frac{4\ln(2)}{\Delta t}\approx 2.772~\Delta t^{-1}.
\eea
The above expressions for the bandwidth include the role
of field fluctuations. However, in our numerical
simulations, we have assumed only Fourier-limited pulses, after convincing ourselves that including the effects of field fluctuations would have only minimal impact on our
chief objective and results. 

\section{Numerical results}

Motivated by recent experiments at FERMI \cite{private}, and in order to focus our discussion on a realistic context, the
parameters used throughout our simulations pertain to the ionization of Neon under radiation of photon energy  $\sim$19 eV. Specifically, neutral Neon in its ground state
configuration $\ket{g}\equiv 2p^6(^1S_0)$, when exposed
to the above photon energy, can be raised to any or both
of two adjacent excited states, namely, 
$(2p^6~^1S)~^1S_0 \to (2p^5~^2P)^2P~(4s^1~^2S)^3P_1 \equiv \ket{a}$ and 
$ (2p^6~^1S)~^1S_0 \to (2p^5~^2P)^2P~(4s^1~^2S)^1P_1\equiv \ket{b}$, 
where 
$\omega_a \simeq 18.82$eV and $\omega_b \simeq 18.91$eV. The respective 
dipole matrix elements are $\mu_{bg}=|\bra{b} \hat{z} \ket{g}| =   0.0995\, \textrm{a.u.}$ and 
$\mu_{ag}=|\bra{a} \hat{z} \ket{g}| =  0.0986\, \textrm{a.u.}$
The ionization energy of Ne is $\hbar\omega_{\rm ion}\approx 20.74$eV.  The extent to which  each of the two
near resonant intermediate states, or both, will dominate 
the REMPI ion yield will depend mainly on the relative detunings  from resonance, the laser bandwidth, as well as the intensity.  
In the recent experiments at FERMI \cite{private}, the FWHM of the FEL pulses was 
$\Delta t\approx 110$fs, which corresponds to a  bandwidth $\Delta\omega_{\min}\approx 16.59$meV. The combined bandwidth $\Delta\omega$ was estimated to about $30$meV and using 
Eqs. (\ref{Dw:eq}) and (\ref{Tc:eq}) we have for the coherence time $T_{\rm c}\approx 109.9$fs. 
This means that fluctuations were barely present in the FEL pulses, and thus throughout this work 
they are ignored by  focusing on Fourier-limited pulses of FWHM  
$\Delta t = 110$fs. Finally, to obtain a better picture of the role of the harmonics, we will consider intensities from $10^{9}{\rm W}/{\rm cm}^2$ to $10^{16}{\rm W}/{\rm cm}^2$; albeit some of the higher intensities  might not be  attainable for the time being at FERMI or other FEL facilities. 
This means that the Rabi frequencies entering our simulations ranged from 6.9$\times 10^{-4}$ to 2.2 rad/fs, which are at least two orders of magnitude smaller than the frequencies of the driven transitions, thus justifying fully the RWA.

The energy of the photoelectrons produced by the absorption of the third harmonic differs from that of the
REMPI by the energy of one photon. As such they can be
easily discriminated from each other; if photoelectron
energies are measured. On the contrary, the discrimination
between photoelectrons due to the second harmonic from those due to REMPI, would require in addition the
measurement of the respective angular distributions. 
Although those two angular distributions would in general be different, as they involve different partial waves, typically it is only at certain angles that the differences may be of sufficient discriminatory resolution;  which depending on the particular atomic system and states involved may not provide a clear separation of the contribution of the 2nd harmonic. Be that as it may, the strategy for optimal detection of the REMPI signal should include energy and angle resolved photoelectron energy spectra, an undertaking requiring a more elaborate experimental set-up, as compared to ion detection only.
That is why, for reasons of
convenience and instrumental simplicity many experiments rely on the measurement of ionic yields only. It is therefore of interest to establish the conditions under which the presence of the harmonics can be safely ignored. 
In this spirit and in order to avoid unnecessary
complexity, we shall consider from here on only the 2nd
harmonic, but our main observations are expected to be valid even when the 3nd harmonic 
is included. This is because the equations of motion depend on the total ionization rate 
at $2\omega$ and $3\omega$ i.e., on $\gamma_g(2\omega)+\gamma_g(3\omega)$, 
and thus by including the third harmonic 
one may expect a slight shift in the intensity at which  features, such as those in Fig. \ref{Yields1:fig}  appear, with no change in the overall behaviour.

In general, the 2nd harmonic is a fraction  $r_2\ll 1$ of the fundamental. Our analysis will be based on the ratio of yields 
\bea
R_{ab} = 
\frac{Y_{\omega}}{Y_{2\omega}^{(g)}} = 
\frac{Y_{\omega}^{(a)}+Y_{\omega}^{(b)}}{Y_{2\omega}^{(g)}}  =R_{a}+R_b
\eea
at the end of the pulse, where $Y_{\omega}$ is the total yield of REMPI produced by the fundamental frequency 
$\omega$. Ideally, one would like to have $R_{ab}\gg 1$ over a broad range of peak intensities so that REMPI
dominates the single-photon ionization yield at $2\omega$.
Thus, in what follows, our objective is two-fold: (a) To
understand the physical processes that affect the dynamics
of the system. (b) To derive rules of thumb that, given
accessible physical parameters, allow us to infer easily
and for a range of intensities the strengh of the REMPI relative to that due to the 2nd  harmonic.

\subsection{Single resonance} 
\begin{figure}
\begin{center}
\includegraphics[scale=0.4]{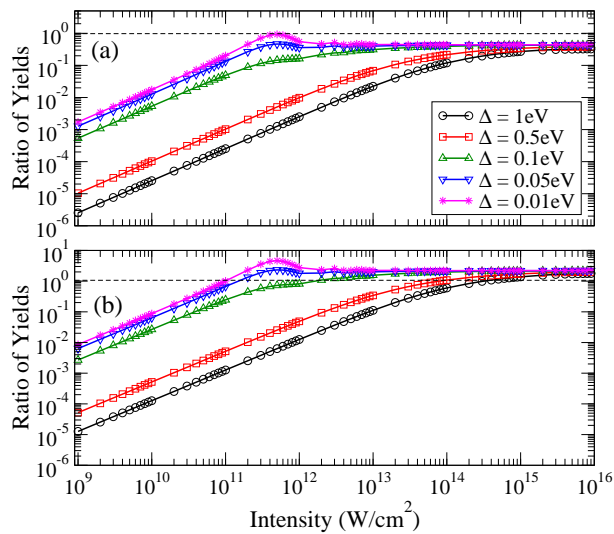}% Here is how to import EPS art
\caption{\label{Yields1:fig} Single resonance. The ratio of yields $R_a$ as a function of the peak intensity 
for various photon-energies and for $r_2 = 1\%$ (a) and $r_2 = 0.2\%$ (b).  Other parameters: $\mu_{bg} = 0$, $r_3 = 0$, Fourier-limited Gaussian pulses of FWHM 110 fs (corresponding bandwidth $\Delta\omega_{\min}\approx 16.59$meV). The dashed lines mark the $R_a=1$ condition.}
\end{center}
\end{figure}

To better understand the interplay of the different
ionization channels in the system, it is instructive to consider first only one of the resonances, by setting $\mu_{bg} = 0$ so that 
$R_{ab} = R_a$. The power dependence of $R_a$,  for different detunings from resonance $\Delta_a$, and for two different 
fractions of the second harmonic is shown in Fig. \ref{Yields1:fig}. We can identify a regime of linear increase 
of the ratio for low intensities, which is followed by some sort of saturation. 

Further insight into the two different regimes can be gained by introducing the pulse area 
\bea
{\cal S}_{a} \equiv \Omega_{a}^{(0)}\int_0^\infty \sqrt{f(t)} dt.
\eea
where $\Omega_{a}^{(0)}$ is the peak Rabi frequency given by Eq. (\ref{RabiEq}), which is proportional to $\sqrt{I^{(0)}}$.  
In the limit of weak excitation i.e., for ${\cal S}\ll 1$ and/or for detunings $\Delta_a \gg \Omega_{a}^{(0)}$,     
the  intermediate resonance can be eliminated. In that case, we have a  two-photon ionization path at frequency $\omega$ competing with the single-photon ionization path
at the harmonic $2\omega$, each of which is describable by
a single rate. 
The two-photon ionization rate is $\gamma_g^{(2)}(\omega;t) = \sigma_g^{(2)}(\omega) F_1^2(t)$, with $\sigma_g^{(2)}(\omega)$ the two-photon cross-section at frequency $\omega$, from which we obtain the yield 
\bea Y_{\omega} &=& 
\int_0^{\infty} dt \gamma_g^{(2)}(\omega;t)\sigma_{gg}(t)   \sim 
\sigma_g^{(2)}(\omega) \int_0^{\infty} dt~ I(t)^2.
\nonumber
\eea
Similarly the yield for the second harmonic is 
\bea
Y_{2\omega}^{(g)} &=& 
\int_0^{\infty} dt \gamma_g(2\omega,t)\sigma_{gg}(t)   \sim \frac{r_2\sigma_g^{(1)}(2\omega)}{2}
  \int_0^{\infty} dt~ I(t)
  \nonumber
\eea 
with $\sigma_g^{(1)}(2\omega)$ the one-photon cross-section 
from $\ket{g}$ at the second harmonic. 
Therefore, as one might have expected, for weak excitation
the ratio scales linearly with the peak intensity   
\bea R_a \sim  \frac{2\sigma_g^{(2)}(\omega) I^{(0)}}{r_2\sigma_g^{(1)}(2\omega)}, 
\eea
as depicted in Fig. \ref{Yields1:fig}.  Again as 
expected, with increasing detuning, the linear regime
extends over a  larger range of intensities.

Formally, for fixed detuning, the condition ${\cal S}_a \sim  1$ marks the end of the linear regime. In that case,
during a Fourier-limited pulse, a significant part of the population is transferred from Ne$[2p^6]$ to Ne[$2p^54s$];
which implies that the intermediate resonance cannot be eliminated.  For exact resonance $\Delta_a=0$,
the dynamics are fully determined by the pulse area.  More precisely, for 
${\cal S}_a = \pi/2$, during the pulse, we have a half  Rabi
oscillation between $\ket{g}$ to $\ket{a}$. Clearly, at
the end of the pulse, only Ne in the excited (resonant)
state is present. For ${\cal S}_a = \pi$ we have a complete 
Rabi oscillation, whereas for ${\cal S}_a > \pi$ many
oscillations take place during the pulse. In other words, the larger the pulse area is, the more oscillations 
take place.

Given that for a fixed fraction $r_2\ll 1$, the highest ratio of yields is obtained on resonance, as in this case REMPI is maximized, we can estimate the lowest possible
threshold peak intensity  $I_{th}$ through the condition ${\cal S}_a = \pi/2$, which marks the end of the linear
regime and the beginning of saturation.  

Although in the case of certain Fourier-limited pulses one  can calculate exactly the integral, in practice this is 
not possible; especially when fluctuations are present. In an attempt to derive a compact rule of thumb for the
threshold intensity we approximate the integral by the FWHM of the pulse obtaining  
\bea
I_{th} = \left [ \frac{10^{15}\pi}{2\times 2.207\times 10^8\times  \mu_{ag} \times \Delta t} \right ]^2
\label{Ith_eq}
\eea 
in ${\rm W}/{\rm cm}^2$, where  Eq. (\ref{RabiEq}) has been used. In this expression the dipole moment $\mu_{ag}$ is in atomic units and 
the FWHM of the pulse $\Delta t$ in femtoseconds.

Increasing the peak intensity further, the ratio of yields exhibits a smooth plateau and approaches some sort of
saturation. This regime is characterized by large Rabi
frequencies and pulse areas  i.e., $\Omega_{a}^{(0)} \gg  \max\{\gamma_a(\omega),\Delta_a\}$ and ${\cal S}_a \gg \pi$, so that many Rabi oscillations take place during the pulse and during the characteristic ionization time $\gamma_a(\omega)^{-1}$. As a result, the ratio of yields can be approximated by 
\bea
R_a^{\infty} \approx \frac{\gamma_a(\omega) \bar{p}_a}{\gamma_g(2\omega) \bar{p}_g},
\eea
where we have introduced the time-averaged populations 
\bea
\bar{p}_j \equiv \frac{1}T \int_0^T dt \sigma_{jj}(t)
\eea
with $T\geq \Delta t$, and $\sum_j \bar{p}_j =1$,  while in all of the above expressions the ionization rates are evaluated at the peak intensity (and thus they are  time independent). 
Using Eqs. (\ref{gamma_j}) and (\ref{F_q})  we obtain for the ratio  
\bea
R_a^{\infty} = \frac{2\sigma_a^{(1)}(\omega)\bar{p}_a}{r_2\sigma_g^{(1)}(2\omega) \bar{p}_g},
\eea
where $\sigma_a^{(1)}(\omega)$ is the one-photon 
cross-section from $\ket{a}$ at the fundamental frequency.
Our simulations demonstrate  that the 
ratio $R_a$ approaches $R_a^{\infty}$ from above in the limit of very large peak intensities. In this limit,  $\bar{p}_j\to 1/2$  and thus $R_a^{\infty} $ depends only
on the ionization cross-sections of the neutral and the excited atom, the order of the harmonic and its relative  intensity with respect to the intensity of the fundamental
frequency. In terms of a physical picture underlying this
behavior, in the regime of parameters where the saturation
has set in, the REMPI has essentially become a single-photon process, whose rate is determined by the ionization
rate of the excited state.

\subsection{Two resonances}

\begin{figure}
\begin{center}
\includegraphics*[scale=0.4]{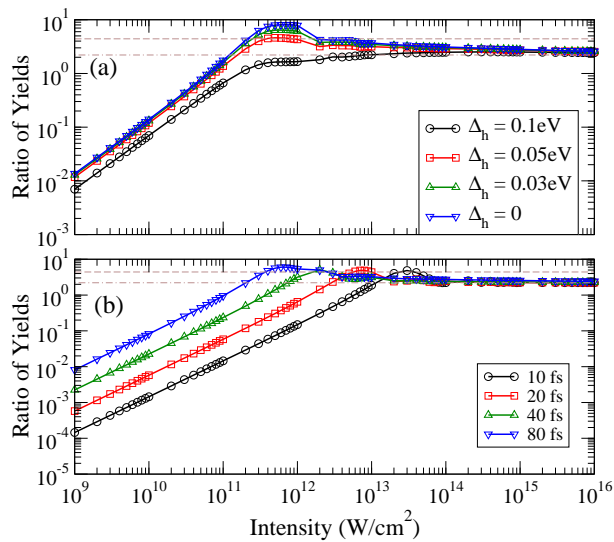}% Here is how to import EPS art
\caption{\label{Yields2:fig} Two resonances. (a) The ratio of yields $R_{ab}$ as a function of the peak intensity 
for various photon-energies  (a) and pulse durations (b).  Other parameters:  (a) $r_2 = 0.2\%$, $r_3 = 0$, Fourier-limited Gaussian pulses of FWHM 110 fs (corresponding bandwidth $\Delta\omega_{\min}\approx 16.59$meV); 
(b) $\Delta_h = 0$  and the other parameters as in (a).  The dashed lines mark the upper and lower bounds
 of the ratio at the plateau, as determined by the equations in the text.}
\end{center}
\end{figure}

The ratio $R_{ab}$ as a function of the peak intensity for the case of two resonances is shown in Fig. \ref{Yields2:fig}(a), 
where the detuning is now measured with respect to frequency $\omega_{h} = (\omega_a+\omega_b)/2$ i.e. 
$\Delta_h = \omega_{h} -\omega_g- \omega$. 
Clearly, the overall behaviour is analogous to the one for single resonance, albeit the ratio 
at a given peak intensity and detuning appears to be somewhat larger than the corresponding ratio for single
resonance,  because of the contribution of the additional resonance to the total REMPI yield. 
One can again identify 
the linear regime for  low intensity, where the intermediate resonances can be eliminated and the 
neutral atom effectively ionizes by single-photon absorption at frequency $2\omega$ and two-photon 
absorption at the fundamental frequency $\omega$.

An unambiguous expression for the threshold intensity, which marks the end of the linear regime, cannot be obtained along the steps outlined above. As depicted in Fig. \ref{Yields2:fig}(a) 
the  highest values of the ratio are not obtained 
on resonance as before, but rather for photon energies half-way between the two resonances. 
Moreover, in any case one of the  resonances will be unavoidably off-resonant, which implies that 
the dynamics of the system depend on the details of the pulse. 
In order to infer  the strength of the REMPI yield relative to the yield of the second harmonic, 
it suffices to obtain a lower bound on the threshold 
intensity. This is possible by considering only one of the resonances i.e., the one with the smallest dipole 
matrix element $\mu_{\eta g}$, and applying Eq. (\ref{Ith_eq}).  

%Given that we are off-resonant we cannot follow the same arguments that allowed as to obtain 
%an expression for the threshold intensity in the case of single resonance. 

In the limit of large intensities (i.e., for $\min_\eta\{\Omega_{\eta}\}\gg \max_\eta\{\gamma_\eta,\Delta_\eta\}$ and large 
pulse areas $\min_\eta\{{\cal S}_\eta\} \gg \pi$), 
the ratio of yields is well approximated by  
\bea
R_{ab}^\infty \approx 
\frac{\gamma_a(\omega) \bar{p}_a +\gamma_b(\omega)\bar{p}_b}{\gamma_g(2\omega)\bar{p}_g}, 
\label{R2inf_2a}
\eea
with ionization rates estimated at the peak intensity. 
Hence,  by means of Eqs. (\ref{gamma_j}) and (\ref{F_q}), Eq. (\ref{R2inf_2a}) can be expressed as 
\bea
R_{ab}^\infty \approx 
2\frac{\sigma_a^{(1)}(\omega) \bar{p}_a +\sigma_b^{(1)}(\omega)\bar{p}_b}{r_g\sigma_g^{(1)}(2\omega)\bar{p}_g}, 
\label{R2inf_2b}
\eea
with $\sum_\eta \bar{p}_\eta = 1$. 

In order to quantify the plateau of the ratio $R_{ab}$, we may consider the upper bound given by 
\bea
4\frac{\max_\eta\{\sigma_\eta^{(1)}(\omega) \bar{p}_\eta\} }{r_2\sigma_g^{(1)}(2\omega) \bar{p}_g}
\eea
and the lower bound  
\bea
2\frac{\min_\eta\{\sigma_\eta^{(1)}(\omega)\} (1-\bar{p}_g) }{r_2\sigma_g^{(1)}(2\omega) \bar{p}_g}.
\eea
Even for the case of two-resonances, for strong driving we expect $\bar{p}_g \to 1/2$ 
and $\max_\eta\{\bar{p}_\eta\}\leq 1/2$ 
obtaining  the following expressions for the upper and lower bounds 
\bea
&&\textrm{Upper bound:~}\, 4\frac{\max_\eta\{\sigma_\eta^{(1)}(\omega)\}}{r_2\sigma_g^{(1)}(2\omega)},
\\
&&\textrm{Lower bound :~}\, 2\frac{\min_\eta\{\sigma_\eta^{(1)}(\omega)\}  }{r_2\sigma_g^{(1)}(2\omega)} 
\eea
which depend only on the cross sections, the order and  the fraction of the harmonic. 
These expressions suffice to give us a quantification of the plateau following the peak of $R_{ab}$. 

\section{Discussion and concluding remarks}

From the detailed analysis and discussion in section IV, it is clear that even a minute amount of 2nd harmonic will
influence the 2-photon yield. The laser intensity dependence of the ratio of the REMPI over the 2nd 
harmonic yield, shown in Figs. \ref{Yields1:fig} and \ref{Yields2:fig}, exhibits a global
behavior. It increases linearly at low intensities, 
reaching eventually a constant value (saturation), which in our formalism is represented by the ratios $R_{a}^{\infty}$ and $R_{ab}^\infty$ for one and two intermediated states, respectively. Obviously, the desired value should be much larger than one, for a broad range of peak intensities so that the REMPI signal dominates over the signal from spurious harmonics. Surprisingly, even for a 2nd harmonic content as low as 0.2\% of the fundamental, that value never exceeds 10. 
Whether this provides sufficient discrimination for a
given experiment will depend on the scope of the experiment.
In the linear regime, the REMPI yield is masked by that of the 2nd harmonic. This feature is independent of the
particular matrix elements and cross sections. 
Their specific values will only influence the intensity at
which the transition to saturation takes place, but not the overall behavior. 
Moreover, we have found that this
overall behavior is independent of whether the photon
energy of the fundamental is in near resonance with one 
or two intermediate states. Actually, all of our 
analytical expressions and numerical results can be
generalized to the presence of more than two intermediate
resonances, without altering the main overall behavior.
Although the participation of more than two intermediate
resonances would be rather unusual, still given the 
occasionally large bandwidth of FEL sources, it might
occur in some experiments.

Anticipating some further questions by the reader, a few
clarifications may be in order here. (a) The usual
presence of intensity fluctuations in FEL beams does not
seem to affect appreciably our chief conclusions, as a result 
of which we have limited our treatment to Fourier-limited pulses. 
(b) In the absence of relevant 
information to the opposite, we have assumed the same
temporal profile for the fundamental and the harmonics. 
Introducing a somewhat different temporal profile (perhaps
narrower ?) for the harmonics, would not alter the overall
behavior. (c) The adoption of a specific set of atomic 
parameters does not entail significant sacrifice of 
generality. After all, typical atomic matrix elements and
cross sections do not differ from each other by orders of
magnitude. Adopting a set of parameters corresponding to
a different atomic system, would only shift the plots in
Figs. \ref{Yields1:fig} and \ref{Yields2:fig} to somewhat different intensities without
affecting the overall behavior \cite{remark}. 
(d) Effects of interaction-volume expansion are expected to be 
present in experiments involving strong radiation, which by necessity is focused. This is an instrumental effect which is apt to affect the observed yields, and as such needs to be taken into consideration in the interpretation of experimental data. It does, however, depend on the particular focusing geometry pertaining to a given experiment, but the relevant theoretical tools are known \cite{LamPRA11}. In the presence of  interaction-volume expansion, the transition to saturation 
is expected to  be smoother, while the intensity at which it takes place may shift. 
The overall behaviour of the yields, however,  will be the same as the one described above.  
  Be that as it may, having results independent of volume effects provides a point of calibration of broader validity, with which volume effects,  pertaining to the particular focusing geometry, must be convoluted

Our closing message would be: Even a small admixture of 
2nd harmonic in REMPI might be deceptively innocuous. For 
a harmonic admixture smaller than the 0.2\% we have 
assumed here, the behavior depicted in Figs. \ref{Yields1:fig} and \ref{Yields2:fig} would
simply occur at somewhat different intensities.

\section{Acknowledgments}
The authors wish to acknowledge informative communications with Giuseppe Sansone concerning experiments at FERMI. This work was supported in part by the European COST Action CM1204 (XLIC).


\begin{thebibliography}{99}
\bibitem{remark} 
For FELs based on planar undulators (producing linearly polarized light),  
2nd harmonic is mainly due to imperfections in the FEL mechanism,  and is weaker than the 3nd  harmonic, which is a consequence of the fact that the electron longitudinal velocity oscillates at one-half of the undulator period.  For helical undulators (producing circularly polarized light), the harmonics are strongly suppressed on-axis, and their presence may be only due to ``imperfections" in the emission mechanism \cite{synchrotron}. Linearly polarized light with suppressed harmonics can be also  produced using a crossed polarized undulator scheme.
\end{thebibliography}
\end{document}